\documentclass[%
reprint,
superscriptaddress,
nofootinbib,
amsmath,amssymb,
aps,
prl,
]{revtex4-2}

\usepackage{graphicx}
\usepackage{dcolumn}
\usepackage{bm}
\usepackage{siunitx}
\usepackage{physics}
\usepackage{soul}

\begin{document}

\title{Resonant Island Trapping in a Fourth-Generation Synchrotron Light Source}

\author{E.~C.~Cortés~García}
\affiliation{CERN, Espl.~des Particules 1, 1217 Geneva, Switzerland}
\affiliation{Deutsches Elektronen-Synchrotron DESY, Notkestr. 85, 22607 Hamburg, Germany }

\author{N. Carmignani}
\affiliation{%
ESRF - The European Synchrotron, 71 Avenue des Martyrs, 38000, Grenoble, France
}%

\author{F. Ewald}
\affiliation{%
ESRF - The European Synchrotron, 71 Avenue des Martyrs, 38000, Grenoble, France
}%


\author{S.~A. Antipov}
\affiliation{%
Deutsches Elektronen-Synchrotron DESY, Notkestr. 85, 22607 Hamburg, Germany 
}%

\author{K. Scheidt}
\affiliation{%
ESRF - The European Synchrotron, 71 Avenue des Martyrs, 38000, Grenoble, France
}%

\author{S. White}
\affiliation{%
ESRF - The European Synchrotron, 71 Avenue des Martyrs, 38000, Grenoble, France
}%

\author{I.~V.~Agapov}%
\affiliation{%
Deutsches Elektronen-Synchrotron DESY, Notkestr. 85, 22607 Hamburg, Germany 
}%

\date{\today}

\begin{abstract}
We report the first direct observation of nonlinear resonance island trapping in a fourth-generation light source with working points far from the excited resonance and examine the nonlinear dynamics and properties of the trapped beam. The discovered dynamics of island trapping may help understand bunch purity and halo formation issues, create additional experimental capabilities for photon science applications, and present means of nonlinear characterization of the machine optics.
\end{abstract}

\maketitle


\textit{\textbf{Introduction.}}~Synchrotron light sources are versatile analytic tools, enabling studies in many areas of science and technology. The latest ---fourth generation \cite{Einfeld:xe5006}--- features highly optimized lattices with high degree of cancellation of nonlinear aberrations, with the hybrid multi-bend achromat lattice concept being the technology of choice for high-energy (6 GeV) machines such as the ESRF EBS  \cite{Raimondi2023}. The residual nonlinearities may cause a number of adverse effects on electron beam dynamics such as halo formation, reduction of dynamic aperture, or beam losses at injection, and need to be controlled. 
Moreover, storage rings are uniquely suitable for general experimental studies of nonlinear oscillations \cite{CERNSPS_FixedLine_NatPhys2024}. For these reasons, exploitation and control of nonlinear beam dynamics has been an important subject in accelerator physics (see e.g. ~\cite{Danilov:2010zz, Kuklev:2019zrf,PhysRevLett.86.3779}).

Among the most exciting nonlinear beam dynamics phenomena is the formation of resonant islands in the phase space of particle motion \cite{Cornacchia:1982pr, Chao:1988xy, Lee:1991si, PhysRevA.46.7942, Wang:1993ib, Ellison:1992uq, Lee:1994vm}. Recently this dynamics found applications in manipulating photon beam properties in light sources~\cite{Ries:IPAC2015-MOPWA021, Holldack2022, OLSSON2021} and slow multi-turn beam extraction from proton synchrotrons~\cite{PRL2002_CappiGiovannozzi, Fraser_CERNSPS_SlowResExt_PRAB2019, PRAB2025-ResExDIV}. In all those cases, betatron frequencies (tunes) of the machine are set close to a resonance condition.

As a result of extensive numerical studies of dynamics of large amplitude particles oscillations in fourth-generation storage rings we found that conditions for trapping of particles into third-order resonance islands can be met without the need to set the optics working point close to the third-order resonance. 
Benefiting from the small natural emittance of EBS,  we could carry out experiments to observe and characterize the trapping dynamics in this new regime. 
In this paper we first describe the dynamics of resonant island trapping at large amplitudes in a fourth-generation light source, and then present the direct experimental observation of island formation, dynamics of the capture process, and the island lifetime measurement at ESRF EBS.

\begin{figure*}[t]
    \centering
    \includegraphics[width=0.9\linewidth]{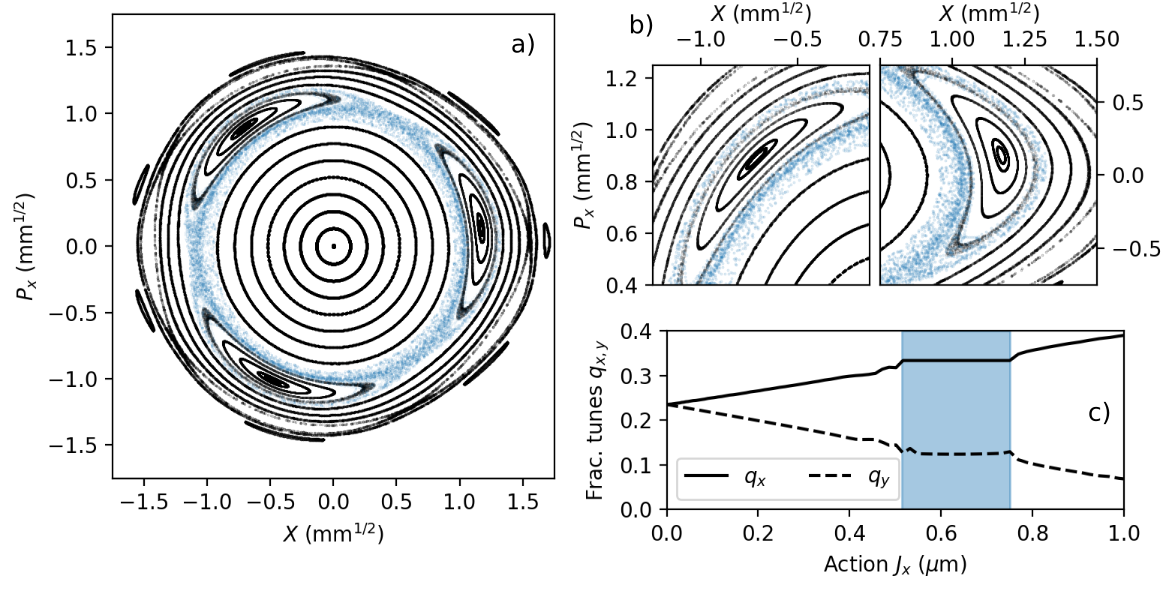}
    \caption{(a) Phase space portrait of the horizontal betatron oscillations. The working point is ($q_x, q_y$) = (0.23, 0.23). The blue band represents the separatrix around the third order resonance islands ($3q_x = 1$). (b) Zoom into the islands: regular motion is found inside the island and a quasi-stochastic layer surrounds them. (c) "Devil's staircase": amplitude dependent fractional tune values for particles along $P_x = 0$ remain constant for a wide range of actions.}
    \label{fig:PhaseSpace-Islands-AmpDetuning}
\end{figure*}

\textit{\textbf{Resonant island trapping process.}}~The transverse beam dynamics in synchrotron light sources without taking radiation damping and excitation into account  can be qualitatively described by the following Hamiltonian:
\begin{align}
    \mathcal{H} = \bar{h}(J_x, J_y) + h^{(r)}(J_x, J_y, \phi_x, \phi_y),
    \label{eq:EffectiveHamiltonian}
\end{align}
where  $(J, \phi)$ are the action-angle variables and the Hamiltonian is split into an (averaged) integrable part and a resonant part which gives rise to the irregular and chaotic motion. The linear part of the averaged Hamiltonian  $\mathcal{H}_0 = -2\pi q_{x,0} J_x$  with $q_{x,0}$ the fractional tune or working point of the machine, represents the linear betatron motion. The action-angle variables are related to the normalized coordinates  as (e.g. for the horizontal plane)
\begin{align}
    \begin{pmatrix}
        X\\P_x
    \end{pmatrix} = \sqrt{2J_x}
    \begin{pmatrix}
       \cos{\phi_x}\\
       \sin{\phi_x}
    \end{pmatrix}.
\end{align}

In practice the dynamics is studied through direct numerical simulations (particle tracking). 
Fig.~\ref{fig:PhaseSpace-Islands-AmpDetuning} shows the transverse phase space of betatron motion of ESRF EBS obtained through tracking.
In this plot, synchrotron oscillations are neglected to  highlight the horizontal dynamics. The dependency of oscillation frequencies on amplitude and crossing of the third-order resonance at large amplitudes is an important feature in the dynamics.

Tracking simulations are carried out using \texttt{pyAT}~\cite{pyat}, incorporating lattice imperfections derived from previous experimental measurements~\cite{liuzzo:ipac2021-tupab048}. At $J_x \approx 0.6$~$\mu$m the system approaches the third order resonance [Fig. 1 (c)], where the resonance part of $\mathcal{H}$ is dominated primarily by 
\begin{equation}
    h^{(r)} = \frac{S}{\sqrt{2}} J_x^{3/2}\sin{(3\phi_x + 3\mu_S)}.
\end{equation}
Then resonance islands emerge in the phase space portrait [Fig. 1 (a-b)].\
Although the amplitude of the resonance driving term $S$ is canceled by design for the nominal optics~\cite{H7BA-ESRF2014}, for high action values the nonlinear optics perturb the cancellation condition.\
The phase $\mu_S$ of the resonance driving term determines the orientation of the islands in phase space.\

High-amplitude particle behavior cannot be fully described by a simple 1D resonance model and an interplay between various resonance conditions may lead to complex dynamics with chaotic behavior and loss of stability.\
Note that the simulation predicts as well a layer overlapping the regular and bounded motion in the resonance islands.\
Here particle tracking simulations are performed in 4D, therefore the crossing of equipotential lines is not prohibited, but rather triggered by parametric resonances described by $h^{(r)}$.\
The actions $J_{\text{FP}}$ at which Hamiltonian fixed-points are found, can be influenced by adjusting the strength of octupoles dedicated to the correction of nonlinear aberrations.\ 
This follows directly from the dependence of amplitude detuning coefficients on the octupole settings.\

To understand the effect of resonance islands on the off-axis beam  injection process, numerical simulations were performed, including the effects of synchrotron radiation and quantum excitation. The beam of emittance \SI{85}{\nano\meter\radian}~\cite{carmignani:ipac2021-mopab051} is injected at normalized position of $X_{\text{inj}} = -$1.16 mm$^{1/2}$ and is damped towards zero as a result of radiation process. 
A significant fraction (80–90\%) of the injected beam damps toward the core ($J_x \rightarrow 0$), while the remainder becomes trapped in resonance islands for at least several hundred milliseconds. The island population is then reduced as a result of quantum diffusion.
The island lifetime is in the range from hundreds of milliseconds to tens of seconds, exhibiting a strong dependency on exact values of the sextupole configuration, which at EBS is tuned to optimize the Touschek lifetime \cite{Liuzzo:2023sud}. 

\begin{figure*}
    \centering
    \includegraphics[width=\linewidth]{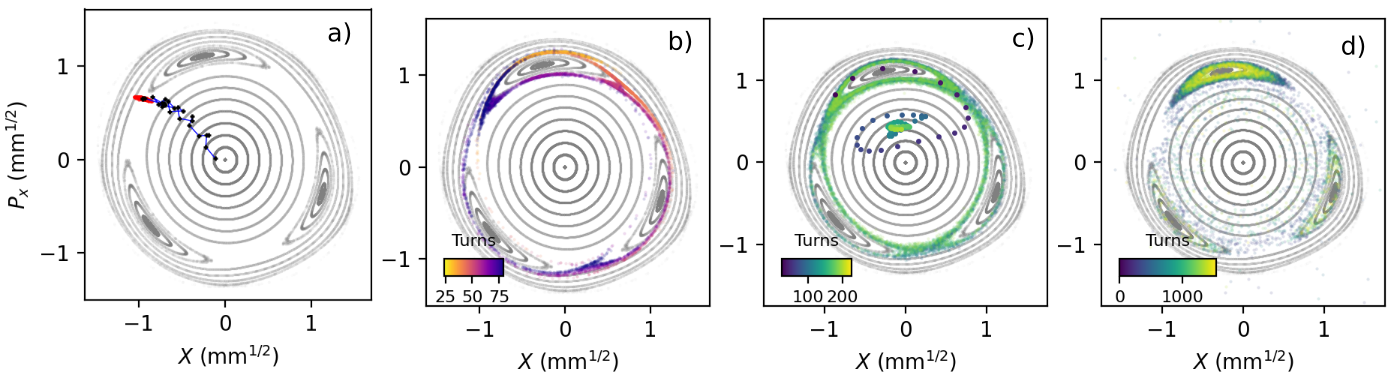}
    \caption{Island capture dynamics simulation. Equipotential levels are included to guide the eye. Data in (b-d) is shown stroboscopically every three turns. The whole data set acquires the same threefold symmetry as the islands. (a) The centroid of the beam over the kicking process is illustrated as black markers. The red blob depicts the resulting distribution after 25 turns during which the kicking takes place. (b) After the beam is brought to the islands, particles exert free betatron oscillations. The particle distribution thereby exhibits strong filamentation due to nonlinear and chromatic aberration effects. (c) A fraction of the beam is trapped in the islands, whereas the rest damps to the center.\ The beam centroid is depicted as dots and follows a spiral (see text). (d) After several turns, the radiation damping dominates and brings the trapped electrons closer to the fixed points determined by the underlying dynamics.}
    \label{fig:CaptureDynamicsSim}
\end{figure*}

The injected beam has larger emittance compared to the stored beam size, and its position in phase space is limited. To characterize the island trapping process better experimentally, it is more effective to apply controlled kicks to the stored beam in order to reach arbitrary locations in phase space and observe the resulting dynamics. Simulation studies were performed in preparation for the experiment. The expected behavior is presented in Fig.~\ref{fig:CaptureDynamicsSim}.\
An ensemble of $10^4$ particles is subjected to kicks over 25 turns, corresponding to the rise time of the kicker pulse [Fig.~\ref{fig:CaptureDynamicsSim} (a)].\
The simulation shows an increase in the RMS action from the equilibrium value of \SI{133}{\pico\meter} to \SI{952}{\pico\meter} as a result of the applied kicks.\
Subsequently, the beam undergoes free betatron oscillations [Fig.~\ref{fig:CaptureDynamicsSim} (b-c)].\
Over the following few hundred turns, strong filamentation develops along the separatrix contour and into the surrounding quasi-stochastic layer. As a result, some particles are redistributed into neighboring resonance islands or along the separatrix.\
Consequently, the beam centroid does not remain near the stable fixed points at the centers of the islands, but instead follows a spiraling trajectory through phase space.\
After this transient capture phase, radiation damping becomes the dominant mechanism, gradually guiding the particles toward the stable fixed points—either within the resonance islands or at the origin. A significant fraction of particles is captured within a single resonance island [Fig.~\ref{fig:CaptureDynamicsSim} (d)]; however, due to diffusion along the separatrix, some particles populate other islands or are damped toward the origin.\
As a result, the beam centroid converges to a position between the origin and the most populated island [Fig.~\ref{fig:CaptureDynamicsSim} (c)].

\begin{table}[h!]
    \caption{Storage ring and beam parameters for the experimental campaign.}
    \label{tab:esrf-ebs-params}
    \centering
\begin{ruledtabular}
\begin{tabular}{lcc}
    Parameter & Symbol & Value\\
    \hline
    Circumference & $C$ & \SI{844}{\meter} \\
    Beam energy   & $E$ & \SI{6}{\giga e\volt}\\
    Beam current   & $I$ & \SI{5}{\milli \ampere}\\
    Fractional tunes & $q_x/q_y$ & 0.23/0.23\\
    Chromaticities & $\xi_x/\xi_y$ & 9/9\\
    Equilibrium emittance & $\epsilon_0$ & \SI{133}{\pico \meter}$\times$rad\\
    Damping times (ms) & ($\tau_x, \tau_y, \tau_z$) & (8.6, 13.2, 8.9) \\
    Eq. momentum spread & $\sigma_p$ & $0.95\times 10^{-3}$\\ 
    \end{tabular}
\end{ruledtabular}\\
\end{table}

\begin{figure}
    \centering
    \includegraphics[width=\linewidth]{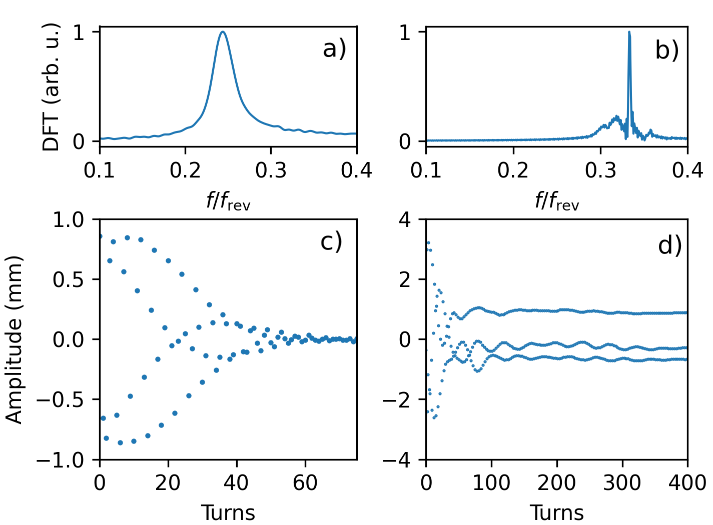}
    \caption{Beam position recorded at one BPM for two different kicker strengths. Lower panels (c, d) depict exemplary recorded BPM signals of the beam centroid oscillations after the electrons have been kicked.\ Top panels (a, b) illustrate their corresponding spectral content.\
    Left panels (a, c) correspond to low kick pulse amplitude, whereas right panels (b, d) show the case where the third order resonance is excited at higher magnet kicker pulse amplitudes.}
    \label{fig:ExampleBPMdata}
\end{figure}

\textit{\textbf{Experiment.}}~Dedicated experiments were performed at EBS with operational machine in the low beam current mode (see Table~\ref{tab:esrf-ebs-params}). The storage ring is filled with a bunch train occupying one-third of the circumference. The electron beam is kicked using four pulsed kicker magnets, and oscillations of its center of mass are recorded with the 320 available beam position monitors (BPMs) in the turn-by-turn fast acquisition mode.
The kickers pulse follow a quarter-sine-wave and has a rise time of \SI{70}{\micro\second}, corresponding to approximately 25 revolutions of the electron beam.\
The pulse amplitude can be set independently for each kicker magnet and thus pre-calculated pulses are set, such that the beam can reach arbitrary points in phase space.
At high pulse amplitudes the beam reaches limits of the dynamic aperture and beam loss is observed.
The phase space is scanned over seven angles in the range $\phi_x \in [0, \pi]$ with steps of $\pi/6$~\SI{}{\radian}.
The amplitude of the magnet kicker pulses is scaled with a single parameter (from 0 to 1) and scanned in ten steps.
The full amplitude corresponds to the expected position of the transverse resonance islands.\
Example BPM data of the centroid oscillations and its corresponding Fourier spectrum are illustrated in Fig.~\ref{fig:ExampleBPMdata}.\
On the left top and bottom panels a typical signal is illustrated.\ 
The beam is kicked and due to chromatic and non-linear tune spread the centroid oscillation signal experiences strong decoherence~\cite{Meller:1987ug}.\
On the right top and bottom panels the case is shown, where the third order resonance is excited, and triple orbits persist.\

\begin{figure}
    \centering
    \includegraphics[width=.9\linewidth]{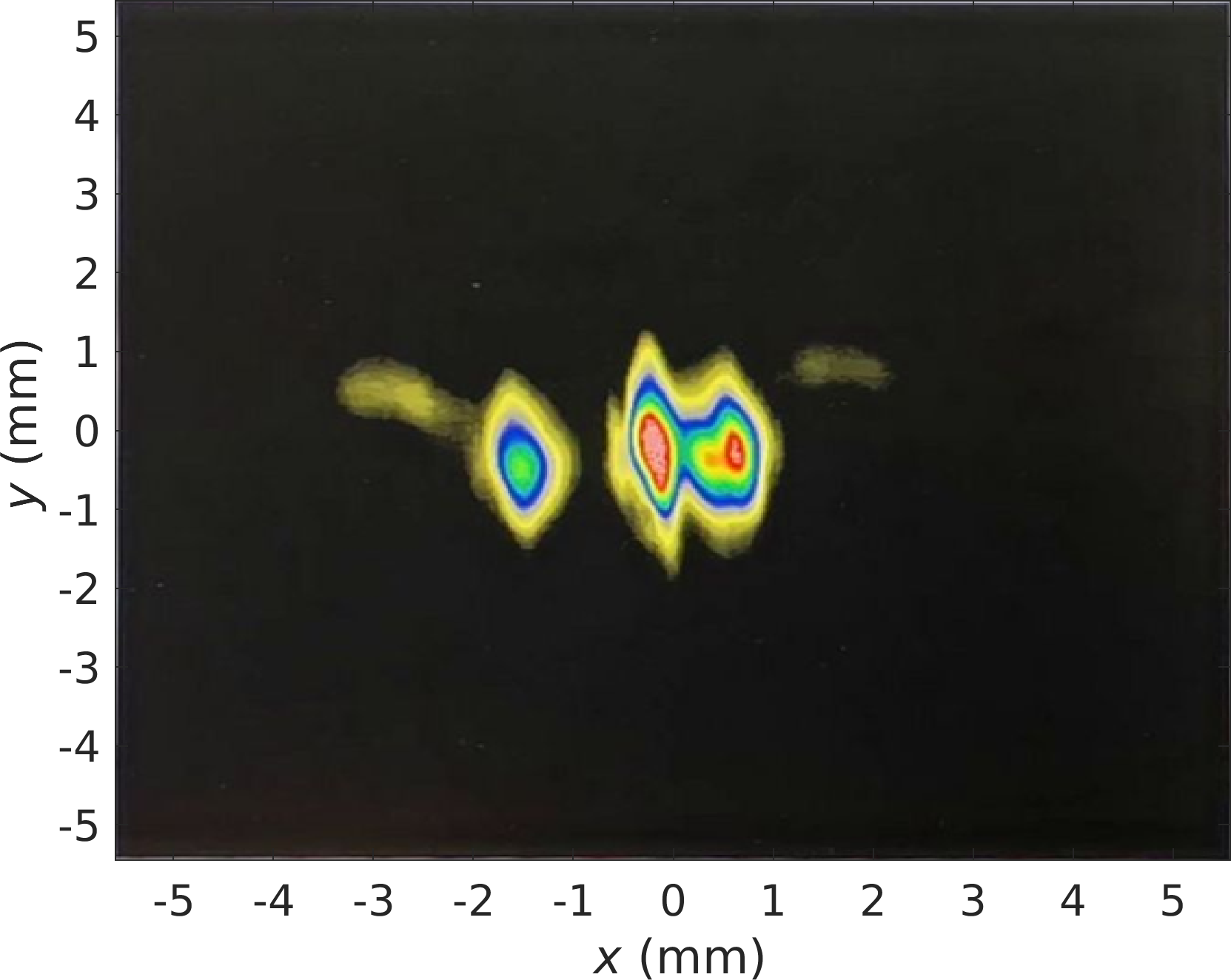}
    \caption{Image of the synchrotron radiation emitted by particles captured in resonance islands. Three distinct islands are visible; the trace near the origin results from the overlap of the stored beam with one of the resonance islands.}
    \label{fig:camera}
\end{figure}

In addition, the radiation emitted by the beam at dedicated imaging diagnostics was recorded, with the results shown in Fig.~\ref{fig:camera}.\
Visible-spectrum synchrotron light from the beam was captured using a camera positioned downstream of a bending magnet.\ 
The experimental setup is described in detail in~\cite{Torino:2019htp}.\
To enhance signal visibility, the acquisition synchronization was delayed by \SI{100}{\milli\second} after the kick, when the islands are fully developed.\
The exposure time was approximately \SI{50}{\milli\second}, and images were acquired at a rate of \SI{15}{\hertz}.\
Due to low capture efficiency (from 3\% to 35\%) the image possesses a high intensity contrast.\

\begin{figure}
    \centering
    \includegraphics[width=1.0\linewidth]{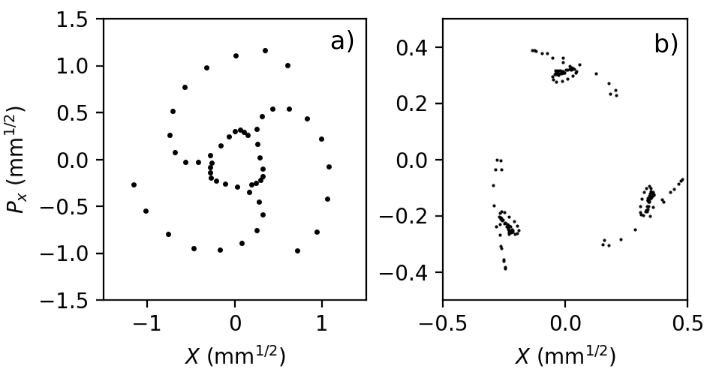}
    \caption{Reconstructed centroid excursion from BPM data. (a) On the left panel the first 51 turns after the kick has taken place are illustrated. (b) The right panel depicts the consecutive 400 recorded turns (starting from the 52nd turn).}
    \label{fig:MeasuredCentroidExcursion}
\end{figure}

To determine the island lifetime, the BPM data is retrieved over a period of \SI{20}{\second} in intervals of approximately \SI{2}{\second} with 200 consecutive turns after the island is populated. 
The amplitude of the BPM signal is proportional to the weighted average of the charge distribution, and the lifetime is determined by fitting the rms of the BPM signal of the three traces with an exponentially decaying function. 
The result is averaged over the data available from the 320 BPMs.
The measured lifetime is $\tau_{\text{island}} = 6.74\text{~s} \pm (0.23_{\text{stat.}} + 0.03_{\text{sys.}})\text{~s}$, three orders of magnitude larger than the damping time due to synchrotron radiation (see Table~\ref{tab:esrf-ebs-params}).\

The data from adjacent BPMs can be used to reconstruct the motion of the center of mass of the beam in phase space \cite{Minty:2003fz}. An example of beam centroid motion reconstructed from 30 BPM pairs during island trapping is shown in  Fig.~\ref{fig:MeasuredCentroidExcursion}.
The top panel shows the first 51 recorded turns after the kicker magnet ramp.
The beam centroid spirals inwards and then converges to a point between the center and the islands fixed points, as predicted by our numerical simulations [Fig.~\ref{fig:CaptureDynamicsSim} (c)]. 

\textit{\textbf{Discussion and outlook.}}~We reported on a new regime of resonant island trapping that occurs in operational optics of low emittance storage rings.\
It results from the interplay of strong nonlinear detuning with low-order resonances.\
The numerical predictions were confirmed experimentally and resonant islands at high oscillation amplitudes with a lifetime of approximately seven seconds were observed through beam position observation and through visible synchrotron radiation.\
These results have several implications on operation of fourth-generation synchrotron light sources. 
They can lead to halo formation and additional beam losses at injection, but, can also be exploited to control the timing structure of photon pulses and create additional experimental capabilities ~\cite{Holldack2020, Holldack2022}.\
Further understanding of dependency of island lifetime on machine nonlinearity can be explored as a tool for diagnostics of nonlinear machine optics.\
Similar phenomena are expected to occur in the planned PETRA IV synchrotron light source~\cite{agapov2024beamdynamicsperformanceproposed}, where a lower emittance beam will provide a higher resolution to study dynamics inside resonance islands, stochastic layers and separatrices. 

\begin{acknowledgments}
\textit{\textbf{Acknowledgments.}}~We want to take the opportunity to warmly thank the PETRA IV and ESRF beam physics groups for their support.\
Our gratitude is extended to Benoit Roche for his support during the experimental campaigns.\
This research was supported in part through the Maxwell computational resources operated at Deutsches Elektronen-Synchrotron DESY, Hamburg, Germany.\
This research was supported in part by the PIER Helmholtz Graduate School Travel Award.
\end{acknowledgments}

\bibliography{main}

\end{document}